\documentclass[%
 aip,
 rsi,
 amsmath,amssymb,
 preprint,
]{revtex4-2}
\pdfoutput=1

\usepackage[utf8]{inputenc}
\usepackage[T1]{fontenc}
\usepackage{graphicx}
\usepackage{dcolumn}
\usepackage{bm}
\usepackage{mathptmx}
\usepackage{etoolbox}
\usepackage{xcolor}
\usepackage{url}
\usepackage[mathlines]{lineno}

\makeatletter
\def\@email#1#2{%
 \endgroup
 \patchcmd{\titleblock@produce}
  {\frontmatter@RRAPformat}
  {\frontmatter@RRAPformat{\produce@RRAP{*#1\href{mailto:#2}{#2}}}\frontmatter@RRAPformat}
  {}{}
}%
\makeatother

\begin{document}


\title[266 nm UV Mach-Zehnder interferometry]{Nanosecond-resolved 266 nm Mach-Zehnder interferometry for electron-density measurements of dense plasmas generated in supercritical fluids}

\author{Kyusang Cho}
\altaffiliation{These authors contributed equally.}
\affiliation{Department of Physics, POSTECH, 77 Cheongam-ro, Nam-gu, Pohang 37673, Republic of Korea}

\author{Juho Lee}
\altaffiliation{These authors contributed equally.}
\affiliation{Department of Physics, POSTECH, 77 Cheongam-ro, Nam-gu, Pohang 37673, Republic of Korea}
\affiliation{Samsung Electronics, Memory Material Technology Team, Republic of Korea}

\author{Gunsu Yun}
\affiliation{Department of Physics, POSTECH, 77 Cheongam-ro, Nam-gu, Pohang 37673, Republic of Korea}
\affiliation{Division of Advanced Nuclear Engineering, POSTECH, 77 Cheongam-ro, Nam-gu, Pohang 37673, Republic of Korea}
\email{gunsu@postech.ac.kr}

\date{\today}

\begin{abstract}
We developed a nanosecond-resolved 266 nm Mach-Zehnder interferometer for electron-density measurements of dense laser-produced plasmas generated in 100-bar supercritical-fluid (SCF) helium. 
A 1064 nm pump pulse was focused into the SCF helium medium, and the plasma-induced phase shift of a 266 nm UV probe beam was recorded using an ICCD-based interferometric imaging system. 
Plasma-arm-only and reference-arm-only images were used to normalize raw interferograms and improve the effective fringe visibility. 
The corrected interferograms were analyzed using a two-dimensional Fourier-transform method to reconstruct phase-shift maps, which were converted into line-integrated electron-density distributions through the plasma dispersion relation. 
Assuming cylindrical symmetry, Abel inversion was applied to the plasma with the largest line-integrated electron density, yielding a maximum local electron density of approximately \(2.5\times10^{18}~\mathrm{cm^{-3}}\). 
The measurement fidelity was evaluated by considering free-free absorption of the probe beam, probe-beam refraction by plasma electron density gradients, and effect of finite-collision-frequency effects on the plasma dielectric response. 
These estimates indicate that the inferred electron density is not altered by more than an order of magnitude under the present experimental conditions. 
The present system demonstrates the applicability of 266 nm UV interferometry to nanosecond-resolved density diagnostics of dense plasmas in high-pressure supercritical fluids.
\end{abstract}

\keywords{UV interferometry; Mach-Zehnder interferometer; laser-produced plasma; dense plasma diagnostics; Abel inversion; supercritical fluids}

\maketitle

\section{Introduction}


High-energy-density physics (HEDP) is the study of matter under extreme conditions, where the energy density, pressure, and temperature are sufficiently large that ordinary descriptions of ideal plasmas are no longer adequate~\cite{REF1}. In its 2003 report, the National Research Council identified the exploration of matter under such extreme conditions as one of the major frontiers of contemporary physics~\cite{REF2}. Under these conditions found in Jovian planetary interiors ~\cite{REF3}, white dwarfs~\cite{REF4}, and solar photosphere ~\cite{REF5, REF6}, most materials are ionized and exist as dense plasmas. Unlike ideal plasmas, dense plasmas can enter non-ideal regimes such as strongly coupled plasmas in which the Coulomb interaction energy becomes larger than the thermal energy, and degenerate plasmas in which the Fermi energy exceeds the thermal energy ~\cite{REF5}. These regimes are commonly characterized by the Coulomb coupling parameter, \begin{equation} 
\Gamma = \frac{e^2}{4\pi\epsilon_0 a k_B T_e},
\end{equation} and the degeneracy parameter,
\begin{equation}
\Theta = \frac{k_B T_e}{E_F},
\end{equation} where \(a\) is the Wigner-Seitz radius, \(k_B\) is the Boltzmann constant, \(T_e\) is the electron temperature, \(\epsilon_0\) is the vacuum permittivity, and \(E_F\) is the Fermi energy. In these regimes, ionization balance, transport, opacity, and equation of state can deviate substantially from those of ideal plasmas~\cite{REF3}.

 Recent advances in high-power lasers, pinch experiment facilities, and accelerators have made it possible to generate high-energy-density matter under controlled laboratory conditions, opening new opportunities for experimental studies of dense plasma physics. 
 Despite these advances, accurate diagnosis of dense plasmas remains challenging. The relevant plasma state is often generated only transiently, typically on time scales shorter than a nanosecond, and the plasma can exhibit high temperature, high density, and significant opacity. These conditions make it difficult to determine the fundamental plasma parameters, such as electron temperature and density. Therefore, time-resolved and quantitatively reliable diagnostics of electron temperature and electron density are essential for experimentally determining plasma states.


In dense plasmas, direct diagnostics using probes are generally not applicable because of the transient and dense plasma conditions.
As a result, non-intrusive optical diagnostics are commonly used, including optical emission spectroscopy, Thomson scattering ~\cite{REF8}, and interferometry~\cite{REF9}. 
Among these methods, interferometry is a representative electron-density diagnostic. 
A probe beam passing through a plasma experiences a phase shift caused by the plasma refractive index, and this phase shift can be converted into a line-integrated electron density using the plasma dispersion relation. 
Interferometric systems can be implemented in various optical geometries, such as Mach-Zehnder~\cite{REF10, REF7}, and Sagnac  configurations, and pulsed laser probes enable density measurements with nanosecond or shorter temporal resolution, enabling the probing of transient events.

For high-density plasma diagnostics, the probe wavelength is a critical parameter. 
The critical density of a probe beam is 
\begin{equation} 
n_c=\frac{\epsilon_0 m_e \omega_{\mathrm{0}}^2}{e^2}= \frac{4\pi^2c^2\epsilon_0 m_e }{e^2\lambda_{\mathrm{0}}^2}\end{equation}
where \(\omega_{\mathrm{0}}\) and \(\lambda_{\mathrm{0}}\) are the probe angular frequency and wavelength, respectively. 
Since \(n_c\propto\lambda_{\mathrm{0}}^{-2}\), shorter wavelengths allow higher-density plasmas to be diagnosed before the probe approaches the cutoff condition~\cite{REF12}. 
A 266 nm UV probe beam has a critical density of approximately \(1.6\times10^{22}~\mathrm{cm^{-3}}\), which is sixteen times larger than that of a 1064 nm laser.
This motivates the use of a 266 nm UV probe beam to measure dense helium laser-produced plasmas in high-pressure supercritical fluids.

In this work, we develop a nanosecond-resolved 266 nm UV Mach-Zehnder interferometer for dense plasmas generated in 100-bar supercritical-fluid (SCF) helium. The diagnostic system combines UV imaging optics, pump-probe synchronization, fringe correction, and Abel inversion. In addition, we evaluate the fidelity of the inferred electron density against probe absorption, probe refraction, and high collision-frequency effect.
\section{266 nm UV Mach-Zehnder interferometer}

\subsection{Experimental apparatus}
\begin{figure*}[t!]
\centering
\includegraphics[width=\linewidth]{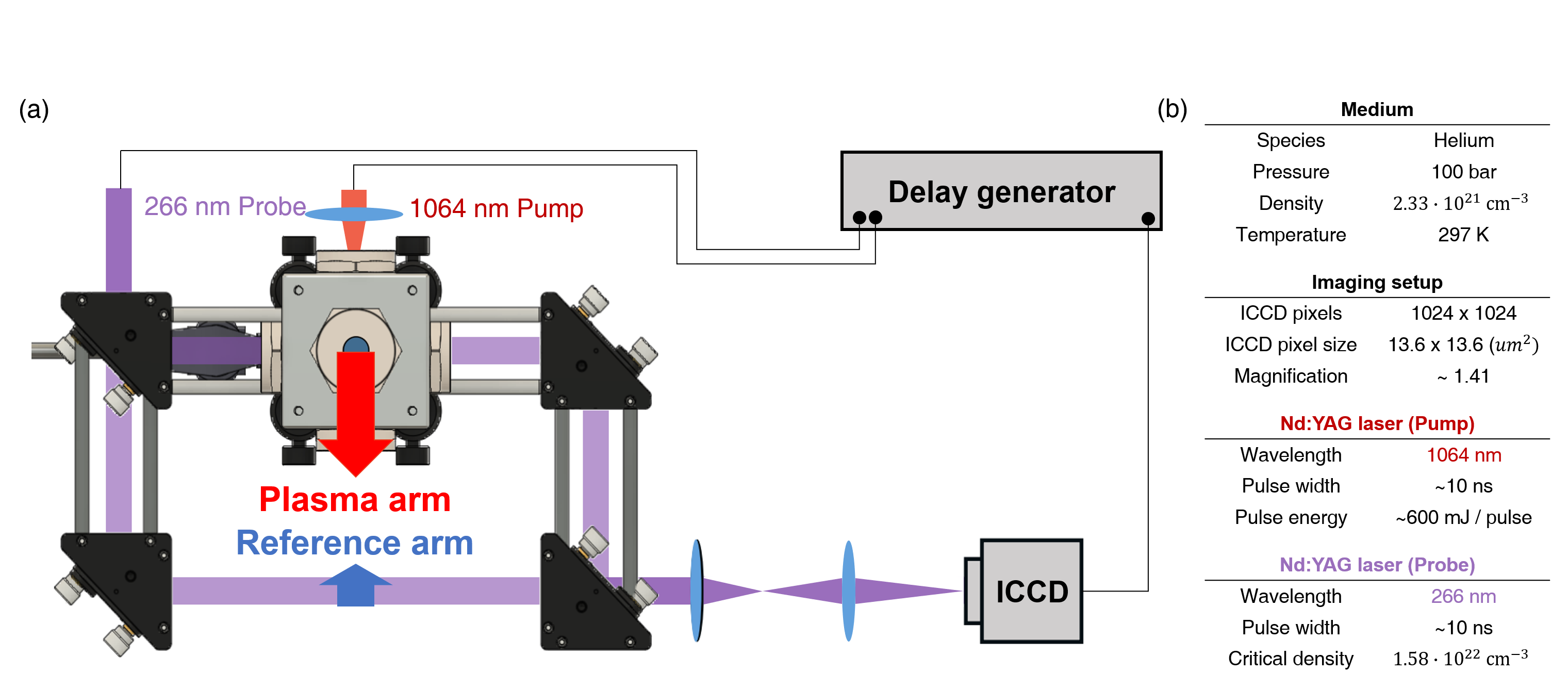}
\caption{
\textbf{Schematic layout of the UV Mach-Zehnder interferometer and experimental conditions.}
\textbf{a.} Schematic of the 266 nm UV Mach-Zehnder interferometer. 
A 1064 nm pump laser and a 266 nm probe laser were synchronized to conduct nanosecond-resolved interferometry. The probe beam was separated into two arms: the plasma arm passed through the plasma, whereas the reference arm propagated through air.
\textbf{b.} Summary of the experimental parameters, including the medium conditions, imaging conditions, and laser parameters. 
}
\label{fig:setup}
\end{figure*}

The experimental procedure consists of two main steps: preparation of the high-pressure SCF medium and laser-plasma generation and pump-probe diagnostics. The medium conditions, imaging conditions, and laser parameters are summarized in figure ~\ref{fig:setup}(b). In the present experiment, 100-bar SCF helium was used as the medium. Helium was chosen because, even if the medium is assumed to be fully ionized without considering physical compression such as radiation compression or shock compression, the corresponding electron density of \(4.66\times10^{21}~\mathrm{cm^{-3}}\) remains below the critical density of the 266 nm UV probe beam. The 5 bar helium gas was pressurized into a high-pressure chamber up to 100 bar using a compressor. After pressurization, the system was left for at least 30 min before the experiment in order to minimize changes in plasma characteristics caused by non-uniformity of the SCF medium ~\cite{REF14}. Once the 100-bar SCF helium was prepared, pump-probe laser plasma generation and diagnostics were performed. A 1064 nm pump laser with a pulse energy of 600 mJ was focused into the chamber using a plano-convex lens (f = 50 mm). At focus, the beam waist was $w_0 = 37~\mu m$, corresponding to a peak focal irradiance of approximately $2.5$~TW/cm$^{2}$. The focused 1064 nm pump pulse generated a dense helium laser-produced plasma.

A 266 nm UV probe beam was synchronized with the 1064 nm pump pulse on a nanosecond timescale using a delay generator. The probe beam was split by a 266 nm beam splitter into two arms: a plasma arm passing through the high-pressure chamber and a reference arm passing through air. The two beams were then recombined by another beam splitter to form interference fringes. The phase difference between the two arms, induced by the refractive-index difference between the plasma and the air, produced a fringe-shift pattern. The resulting interference pattern was imaged onto an ICCD camera at a magnification of approximately 1.43 with a 3 ns ICCD gate width using a UV-coated two-lens imaging system. We used a two-lens imaging system consisting of 100-mm and 50-mm focal-length lenses for better image quality. A 266 nm band-pass filter and an iris, which are omitted from figure~\ref{fig:setup}(a) for simplicity, were placed in the imaging path to suppress stray light and improve image quality. For the fringe-correction procedure described in section 2.2, additional images were acquired under the same spatial and temporal conditions by allowing the probe beam to pass only through the plasma arm or only through the reference arm. These images were used to correct the fringe patterns of the interferograms before phase reconstruction.

In addition to interferometry, plasma imaging was performed to compare the electron-density distribution with the optical appearance of the plasma. For plasma imaging, the 266 nm UV optics shown in figure~\ref{fig:setup}(a) were replaced with silver-coated mirrors, while the same imaging geometry, magnification, and delay-generator timing scheme were maintained. However, the spatial position of the plasma image was not identical to that in the interferometric configuration because of slight alignment changes.

\subsection{Fringe correction}
\begin{figure*}[t!]
\centering
\includegraphics[width=\linewidth]{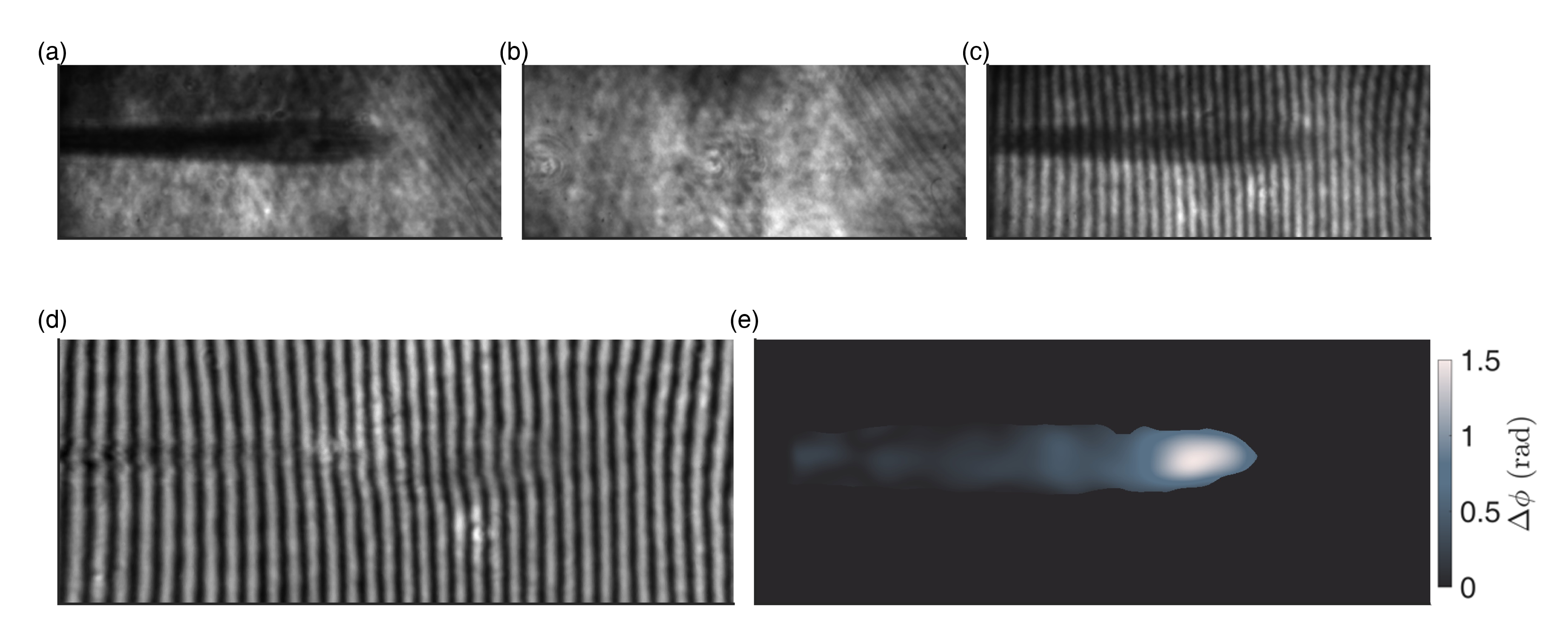}
\caption{
\textbf{Example of fringe correction procedure.}
All images were captured at 10 ns after plasma generation.
\textbf{a.} Plasma-arm-only image.
\textbf{b.} Reference-arm-only image.
These two images were used to correct the raw interferogram fringe patterns.
\textbf{c.} Raw interferogram. 
\textbf{d.} Corrected interferogram obtained by normalizing the raw interferogram using the images shown in \textbf{a} and \textbf{b}. 
\textbf{e.} Two-dimensional phase-shift map in radians reconstructed from \textbf{d} using the two-dimensional FFT method.
}
\label{fig:fringe}
\end{figure*}


Nanosecond-resolved raw interferograms were obtained using the 266 nm UV Mach-Zehnder interferometer described in section 2.1. 
The raw interferogram intensity is given by
\begin{equation}
I_{\mathrm{raw}}
=
I_p
+
I_r
+
2\sqrt{I_pI_r}
\cos\left(\Delta\phi\right),
\label{eq:raw_interferogram}
\end{equation}
where \(I_p\) and \(I_r\) are the intensities of the plasma-arm only and reference-arm only, respectively, and \(\Delta\phi\) is the phase difference. 
The fringe visibility is maximized when \(I_p=I_r\), but this condition is not perfectly satisfied in the experiment because of various optical losses such as reflection and scattering at the high-pressure SCF medium. To improve the fringe contrast and remove the spatial intensity imbalance between the two arms, the raw interferogram was normalized using the plasma-arm-only image in figure~\ref{fig:fringe}(a) and the reference-arm-only image in figure~\ref{fig:fringe}(b):
\begin{equation}
I_{\mathrm{corr}}
=
\frac{
I_{\mathrm{raw}}-I_p-I_r
}{
2\sqrt{I_pI_r}
}
=
\cos\left(\Delta\phi\right).
\label{eq:fringe_correction}
\end{equation}
The corrected interferogram is shown in figure~\ref{fig:fringe}(d). 
The shadow-like intensity reduction observed in the raw interferogram is largely suppressed after the correction, and the fringe shift in the plasma region becomes clearer.

The corrected interferogram was then processed using a 2D FFT method. After reconstructing the phase, the reference phase measured without plasma was subtracted to obtain the plasma-induced phase-shift map: 
\begin{equation}
\Delta\phi
=
\phi_{\mathrm{plasma}}
-
\phi_{\mathrm{ref}}.
\label{eq:phase_shift}
\end{equation}
The resulting two-dimensional phase-shift map is shown in figure~\ref{fig:fringe}(e), which is later converted into a line-integrated electron-density distribution.

\subsection{Electron-density reconstruction and Abel inversion}

\begin{figure*}[t!]
\centering
\includegraphics[width=\linewidth]{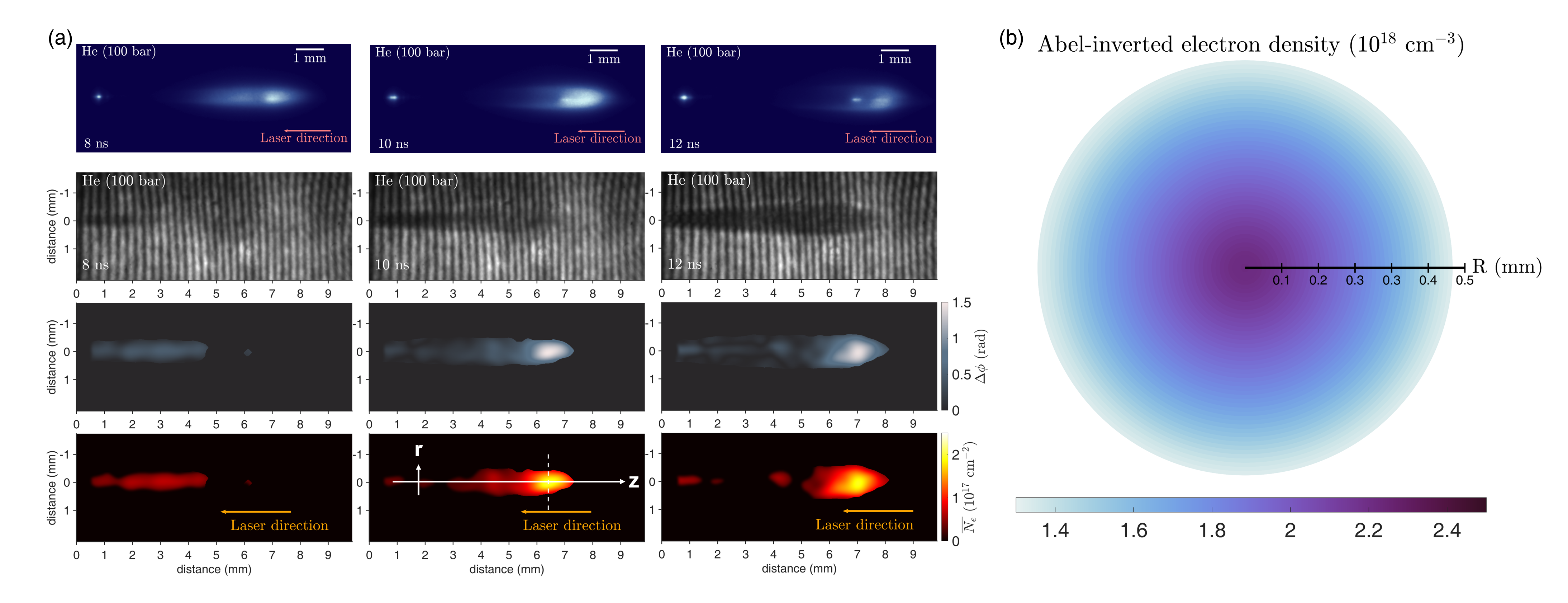}
\caption{
\textbf{Plasma imaging, interferometry, line-integrated electron-density maps, and Abel-inverted electron-density profile of dense helium plasma.}
\textbf{a.} Diagnostic results obtained at 8 ns, 10 ns, and 12 ns after plasma generation. 
From top to bottom, the rows show plasma images, raw interferograms, reconstructed two-dimensional phase-shift maps, and two-dimensional line-integrated electron-density maps. 
The plasma imaging and interferometry measurements were performed at the same pump-probe delays and at the same imaging magnification, but the apparent plasma positions in the imaging and interferometry data are not spatially identical. 
The pump laser propagates from right to left in the images. 
The generated plasma expands backward with respect to the pump propagation direction, toward the laser-incident side. The phase maps show phase shifts in radians. The line-integrated electron-density maps were calculated from the phase maps and are shown in units of \(\mathrm{cm^{-2}}\).
\textbf{b.} Abel-inverted local electron-density profile obtained from the column with the maximum line-integrated electron density in the 10 ns data shown in \textbf{a}. 
The selected column is indicated by the white dashed line in the 10 ns line-integrated electron-density map. 
Assuming cylindrical symmetry about the \(z\)-axis, the line-integrated electron density was converted into a radial electron-density profile. 
}
\label{fig:result}
\end{figure*}

As described in section~2.2, a clear two-dimensional phase-shift map was reconstructed by applying fringe correction and a two-dimensional FFT method to the raw interferograms. 
In 266 nm UV Mach-Zehnder interferometry, the phase difference between the two arms arises because the reference-arm probe propagates only through air, whereas the plasma-arm probe traverses the plasma over a finite path length \(L_{p}\). 
This difference in optical path produces the fringe shift observed in the interferogram.

The relative permittivity of the plasma is generally given by ~\cite{REF10}
\begin{equation}
N^2
=
1-
\frac{\omega_p^2}{\omega_{\mathrm{0}}^2}
\frac{1}{1+i\nu_c/\omega_{\mathrm{0}}},
\label{eq:plasma_refractive_index_general}
\end{equation}
where  \(N\) is the plasma refractive index, \(\omega_p\) is the electron plasma frequency, \(\omega_{\mathrm{0}}\) is the angular frequency of the probe beam, and \(\nu_c\) is the effective collision frequency. 
If the effective collision frequency is much smaller than the probe angular frequency, i.e. \(\nu_c \ll \omega_{\mathrm{0}}\), the plasma relative permittivity can be approximated as
\begin{equation}
N^2
=
\frac{k^2c^2}{\omega_{\mathrm{0}}^2}
\simeq
1-\frac{\omega_p^2}{\omega_{\mathrm{0}}^2}
=
1-\frac{n_e}{n_c},
\label{eq:plasma_refractive_index_approx}
\end{equation}
where \(k\) is the wavenumber, \(n_e\) is the electron density, and \(n_c\) is the critical density of the probe beam.

The phase difference \(\Delta\phi\) observed in the interferogram originates from the refractive-index difference between the plasma and air, and can be written as
\begin{equation}
\Delta\phi
=
\int_{0}^{L_p}
\left(N-1\right)
\frac{\omega_{\mathrm{0}}}{c}\,dl.
\label{eq:phase_shift_general}
\end{equation}
where \(L_p\) is the effective path length of the probe beam through the plasma.
If the electron density is much smaller than the probe critical density, \(n_e \ll n_c\), equation~\eqref{eq:phase_shift_general} can be expanded to give
\begin{equation}
\Delta\phi
\simeq
-\int_{0}^{L_p}
\frac{n_e}{2n_c}
\frac{\omega_{\mathrm{0}}}{c}\,dl
.
\label{eq:phase_shift_lineintegrated}
\end{equation}
Therefore, the line-integrated electron density can be obtained directly from the reconstructed phase-shift map as
\begin{equation}
\int_{0}^{L_p} n_e\,dl
=
-\frac{\lambda_{\mathrm{0}}n_c}{\pi}\Delta\phi.
\label{eq:lineintegrated_density}
\end{equation}
Using this relation, the two-dimensional phase-shift maps shown in the third row of figure~\ref{fig:result}(a) were converted into the two-dimensional line-integrated electron-density maps shown in the fourth row.

Previous studies of dense laser-produced plasmas in 100-bar SCF helium have shown that the plasma reaches its highest brightness, temperature, and expansion length at approximately 10 ns after plasma generation~\cite{REF15}. 
Consistent with those observations, the UV interferometry measurements in the present work showed that the line-integrated electron density reached its maximum at 10 ns, with a peak value of approximately \((2\sim2.5)\times10^{17}~\mathrm{cm^{-2}}\). 
From the corresponding plasma image, the plasma diameter in the selected column was estimated to be less than \(1~\mathrm{mm}\), indicating that the characteristic electron density is on the order of \(10^{18}~\mathrm{cm^{-3}}\).

To estimate the maximum local electron density of the plasma, Abel inversion was applied to the column showing the highest line-integrated electron density at 10 ns. 
Under the assumption of cylindrical symmetry about the \(z\)-axis, the line-integrated electron density \(\bar N_{\mathrm{e}}\) and the local electron density \(n_e(r)\) are related by the Abel transform ~\cite{REF16, REF17},
\begin{equation}
\bar N_{\mathrm{e}}
=
2\int_{y}^{R}
\frac{n_e(r)\,r}{\sqrt{r^2-y^2}}\,dr,
\label{eq:abel_forward}
\end{equation}
and the corresponding inverse Abel transform is
\begin{equation}
n_e(r)
=
-\frac{1}{\pi}
\int_{r}^{R}
\frac{d\bar N_{\mathrm{e}}(y)/dy}{\sqrt{y^2-r^2}}\,dy.
\label{eq:abel_inverse}
\end{equation}
The plasma diameter used in the inversion was estimated from the shadow width of the selected column, and the result is shown in figure~\ref{fig:result}(b). 
The reconstructed local electron density is highest at the plasma center and decreases with increasing radial distance \(r\). 
The maximum local electron density at the center was estimated to be approximately \(2.5\times10^{18}~\mathrm{cm^{-3}}\).
\section{Fidelity of the UV interferometry results}
\begin{figure*}[t!]
\centering
\includegraphics[width=\linewidth]{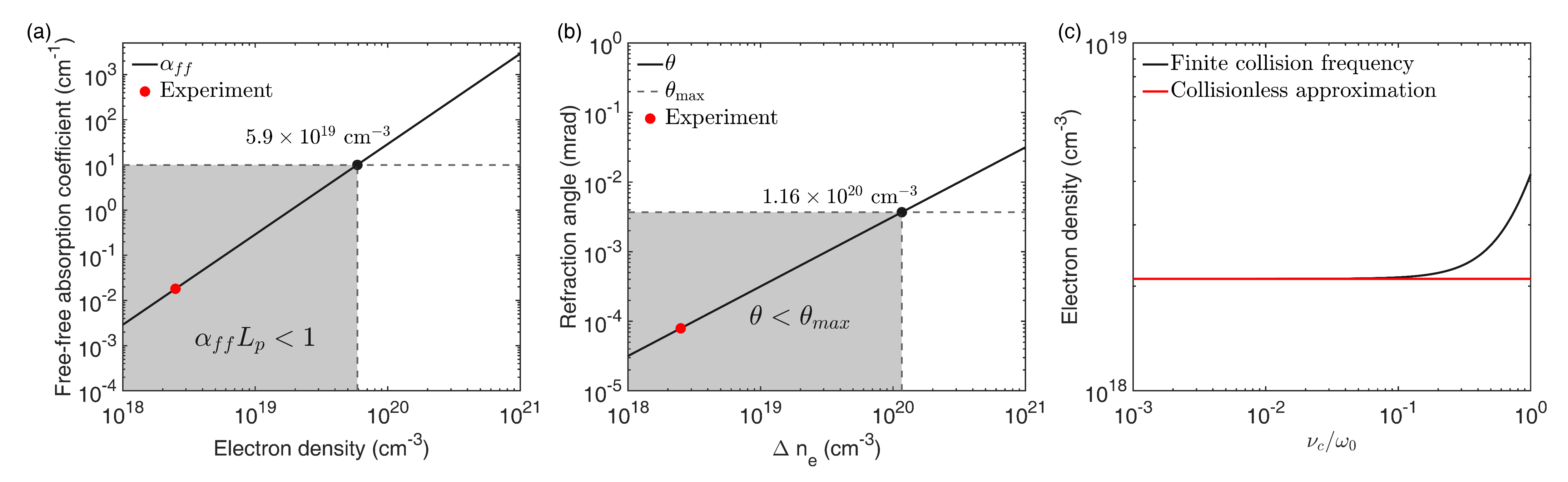}
\caption{
\textbf{Fidelity assessment of the UV Mach-Zehnder interferometry results.}
\textbf{a.} Probe-beam absorption criterion calculated from the optical-depth condition \(\alpha_{ff} L_p<1\), where \(\alpha_{ff}\) is the free-free absorption coefficient. 
The gray shaded region represents the electron-density range satisfying this condition for \(k_BT_e=0.7~\mathrm{eV}\) and the estimated plasma path length \(L_p\). The red dot represents the experimental condition with \(n_e = 2.5\times10^{18}~\mathrm{cm^{-3}}\). 
\textbf{b.} Refraction criterion for the probe beam. \(\theta\) and \(\theta_{max}\) are refraction angle and the maximum allowed refraction angle determined by the interferometer geometry, respectively.
The gray shaded region represents the transverse electron-density-difference range satisfying the condition of \(\theta < \theta_{max}\) and the estimated plasma path length \(L_p\). The red dot represents the experimental condition.
\textbf{c.} Collision-frequency correction to the electron density inferred from the measured fringe shift. 
The red and black solid lines represent the electron density under collisionless approximation and with finite collision frequency included, respectively. 
}
\label{fig:fidelity}
\end{figure*}

As described in section~2.3, interferometry determines the electron density by quantitatively measuring the phase difference between a probe beam that passes through the plasma and a reference beam that passes through the air. 
The measured phase difference is then converted into a line-integrated electron density using the plasma dispersion relation. 
This interpretation assumes that the measured fringe shift is mainly caused by the real part of the plasma refractive-index change, that the probe beam propagates along a well-defined optical path through the plasma, and that sufficient fringe visibility is maintained for reliable phase reconstruction.
In dense laser-produced plasmas, however, several effects can compromise these assumptions. 
First, absorption of the probe beam by plasma can reduce the transmitted intensity and fringe visibility. 
Second, steep electron-density gradients can refract the probe beam.
In such a case, the measured phase difference may no longer represent the intended line integral of the electron density along a straight optical path, resulting in low fringe contrast. 
Third, in dense plasmas, the effective collision frequency can become non-negligible compared with the probe angular frequency. 
This modifies the plasma relative permittivity and can lead to an underestimation of the electron density if the collisionless approximation is used.
Therefore, to assess the fidelity of the electron density inferred from the 266 nm UV Mach-Zehnder interferometer, we quantitatively evaluated the following three effects: probe-beam absorption, probe-beam refraction, and the influence of high collision-frequency in dense plasmas. 
The corresponding analyses are presented in sections~3.1--3.3.

\subsection{Probe-beam absorption}

In dense laser-produced plasmas with electron densities on the order of 
\(10^{18}~\mathrm{cm^{-3}}\) or higher, attenuation of the probe beam can be affected by free-free absorption, also referred to as inverse bremsstrahlung ~\cite{REF12, REF18}.
If the probe beam is strongly absorbed in the plasma arm, the transmitted probe intensity decreases and the fringe visibility is reduced in the plasma region. 
This can degrade the accuracy of the Fourier-based phase reconstruction and may lead to an unreliable electron-density estimate. 
Therefore, we first evaluated whether free-free absorption of the 266 nm probe beam is significant under the present experimental conditions.

The free-free absorption coefficient was estimated using the following expression ~\cite{REF19}:
\begin{equation}
\alpha_{\mathrm{ff}}
\simeq
2.44 \times 10^{-37}
\frac{
 Z^2 n_e n_i
}{
\sqrt{k_{\mathrm B}T_e}\,(h\nu)^3
}
\left[
1-\exp\left(
-\frac{h\nu}{k_{\mathrm B}T_e}
\right)
\right]
\quad \mathrm{cm^{-1}},
\label{eq:free_free_absorption}
\end{equation}
where  \(Z\) is the ion charge state, \(n_e\) and \(n_i\) are the electron and ion densities in \(\mathrm{cm^{-3}}\), respectively, \(k_{\mathrm B}T_e\) is the electron temperature in eV, and \(h\nu\) is the probe photon energy in eV. 
For the 266 nm probe beam used in this experiment, \(h\nu\) is approximately \(4.66~\mathrm{eV}\). 
In this estimate, the attenuation of the probe beam was assumed to be caused only by free-free absorption. 
The helium plasma was also assumed to be singly ionized, so \(Z = 1\) and \(n_i = n_e\).

The optical depth of the probe beam through the plasma is given by
\begin{equation}
\tau_{\mathrm{ff}} = \alpha_{\mathrm{ff}} L_p,
\label{eq:optical_depth}
\end{equation}
where \(L_p\) is the effective path length of the probe beam through the plasma. 
Here, \(L_p\) was taken to be \(1~\mathrm{mm}\), corresponding to the approximate plasma diameter obtained from plasma imaging. 
As a practical absorption criterion for reliable fringe analysis, we required
\begin{equation}
\alpha_{\mathrm{ff}} L_p < 1.
\label{eq:absorption_criterion}
\end{equation}
This condition corresponds to a transmitted intensity larger than \(e^{-1}\) when free-free absorption is the only attenuation mechanism.

Figure~\ref{fig:fidelity}(a) shows the calculated free-free absorption coefficient for a helium plasma at \(k_{\mathrm B}T_e = 0.7~\mathrm{eV}\), which is the electron temperature estimated from previous measurements of laser-produced plasma in 100-bar SCF helium under the same experimental conditions ~\cite{REF15}. 
The electron density was varied from \(10^{18}\) to \(10^{21}~\mathrm{cm^{-3}}\). 
For \(h\nu=4.66~\mathrm{eV}\) and \(L_p=1~\mathrm{mm}\), the optical-depth condition \(\alpha_{\mathrm{ff}}L_p<1\) is satisfied up to an electron density of approximately \(5.9\times10^{19}~\mathrm{cm^{-3}}\). 
The gray shaded region in figure~\ref{fig:fidelity}(a) indicates the density range satisfying this absorption criterion.

The maximum electron density obtained in the present helium plasma experiment was on the order of \(10^{18}~\mathrm{cm^{-3}}\), which is well inside the region satisfying \(\alpha_{\mathrm{ff}}L_p<1\). 
For example, at \(n_e = 2.5\times10^{18}~\mathrm{cm^{-3}}\), the estimated optical depth is approximately \(1.8\times10^{-3}\). 
Therefore, free-free absorption of the 266 nm probe beam is not expected to significantly reduce the fringe visibility or invalidate the interferometry under the present experimental conditions.

\subsection{Probe-beam refraction}

When a probe beam propagates through a medium with a steep density gradient, the probe ray can be refracted from its original propagation direction. 
Such probe-beam refraction can modify the effective optical path through the plasma and reduce the fringe visibility at the imaging plane ~\cite{REF20}.
The refraction of the probe beam can be described in terms of the local wavenumber: 
\begin{equation}
k(x)
=
\frac{\omega_0}{c}
\left(
1-\frac{\omega_{p}^2}{\omega_0^2}
\right)^{1/2}
\simeq
\frac{\omega_0}{c}
\left[
1-\frac{1}{2}
\frac{\omega_{p}^2}{\omega_0^2}
\right],
\qquad
\left(\omega_0 \gg \omega_{p}\right).
\label{eq:local_wave_number}
\end{equation}
where \(\omega_0\) is the probe angular frequency.
When a transverse electron-density gradient is present, \(\omega_{p}\) varies across the beam between two transverse positions, \(x\) and \(x+\Delta x\).
This difference is estimated as~\cite{REF20}
\begin{equation}
\begin{aligned}
\Delta k 
&= k_x - k_{x+\Delta x} \\
&\simeq 
\frac{\omega_{p}^2(x)-\omega_{p}^2(x+\Delta x)}
{2c\omega_0} \\
&=
\frac{\lambda_0 e^2\left[n_e(x)-n_e(x+\Delta x)\right]}
{4\pi c^2 m_e\epsilon_0} \\
&=
\frac{\lambda_0 e^2}{4\pi c^2 m_e\epsilon_0}\Delta n_e,
\end{aligned}
\label{eq:wave_number_shift}
\end{equation}
where \(\lambda_0\) is the probe wavelength, \(\omega_{p}\) is the electron plasma frequency, and \(\Delta n_e \equiv n_e(x) - n_e(x+\Delta x)\) is the transverse electron-density variation.

If the probe beam passes through a plasma over a path length \(L_{\mathrm{p}}\) and an transverse electron-density variation \(\Delta  n_e\), the corresponding refraction angle can be written as
\begin{equation}
\theta
\simeq
L_{\mathrm{p}}
\Delta n_e
\frac{\lambda_0^2 e^2}
{8\pi^2 c^2 m_e\epsilon_0}.
\label{eq:refraction_angle_density_gradient}
\end{equation}
where \(e\) is the elementary charge, \(c\) is the speed of light, \(m_e\) is the electron mass, and \(\epsilon_0\) is the vacuum permittivity, respectively.
This equation shows that the refraction angle increases linearly with the transverse electron-density variation. 
The dense helium laser-produced plasma observed in the present experiment has a sharp boundary, as seen in the plasma images and phase-shift maps in figure~\ref{fig:result}(a). 
Therefore, a steep electron-density gradient can be expected near the plasma edge, resulting in probe-beam refraction. 

To estimate the maximum allowed refraction angle in our experimental setup, we followed the geometrical analysis used in previous studies~\cite{REF20}. 
If the probe beam is refracted by a small angle \(\theta\), the resulting optical-path difference at the imaging plane can be approximated as
\begin{equation}
\delta
=
D'\sin\theta
\simeq
D'\theta
=
D\theta
\frac{f}{L_{\mathrm{obj}}-f},
\label{eq:path_difference_refraction}
\end{equation}
where \(D\) is the initial probe-beam diameter, \(D'\) is the beam diameter at the image plane, \(f\) is the focal length of the imaging lens, and \(L_{\mathrm{obj}}\) is the distance from the refractive medium to the lens. 
The phase difference associated with this additional optical-path difference is
\begin{equation}
\Delta\Phi
=
\frac{2\pi}{\lambda_0}\delta
\simeq
\frac{2\pi}{\lambda_0}
D\theta
\frac{f}{L_{\mathrm{obj}}-f}.
\label{eq:phase_shift_refraction}
\end{equation}
As a criterion for reliable interferometric analysis, we required the absolute value of the refraction-induced phase difference to be smaller than \(\pi/8\), following the criterion used in the reference study. 
This condition gives the maximum allowed refraction angle as
\begin{equation}
\theta_{\mathrm{max}}
\simeq
\frac{\lambda_0}{16D}
\frac{L_{\mathrm{obj}}-f}{f}.
\label{eq:allowable_refraction_angle}
\end{equation}

For a 266 nm probe beam, \(\omega_0 = 7.1\times10^{15}~\mathrm{rad/s}\), initial probe-beam diameter of \(D=9~\mathrm{mm}\), an imaging-lens focal length of 100 mm, and \(L_{\mathrm{obj}}\) is approximately 300 mm.
Using the above parameters, equation~\eqref{eq:allowable_refraction_angle} gives
\begin{equation}
\theta_{\mathrm{max}} = 
3.69\times10^{-3}~\mathrm{mrad}.
\label{eq:theta_max_value}
\end{equation}
This value is indicated by the horizontal gray dotted line in figure~\ref{fig:fidelity}(b). 
The calculated refraction angle increases linearly with the transverse electron-density gradient, as shown by the black solid line in figure~\ref{fig:fidelity}(b). 
By comparing the calculated refraction angle with \(\theta_{\mathrm{max}}\), we find that sufficient fringe contrast can be maintained when the transverse electron-density variation is below approximately \(\Delta n_e<1.16\times10^{20}~\mathrm{cm^{-3}}\) under the present optical geometry. The peak electron density from experiment, on the order of \(10^{18}~\mathrm{cm^{-3}}\), is well below this limit.

\subsection{Electron-density underestimation by finite collision frequency}

In section~2.3, the electron density was inferred using the collisionless plasma dispersion relation under the assumption that the electron density is much smaller than the critical density of the 266 nm probe beam. 
This approximation is generally valid when the collision frequency is negligible compared with the probe angular frequency. 
However, in dense laser-produced plasmas, the effective collision frequency can become non-negligible~\cite{REF21}. 
In such a case, the collisionless plasma refractive index may not be adequate for converting the measured phase shift into electron density. 
Therefore, we evaluated how much the inferred electron density can be underestimated when the effective collision frequency is ignored.

When an effective collision frequency is included, the complex relative permittivity of the plasma can be written as
\begin{equation}
\tilde{\epsilon}
=
1-
\frac{\omega_p^2}
{\omega_{\mathrm{0}}\left(\omega_{\mathrm{0}}+i\nu_c\right)}
=
\epsilon_r + i\epsilon_i,
\label{eq:complex_permittivity}
\end{equation}
where \(\omega_{\mathrm{0}}\) is the angular frequency of the probe beam, \(\nu_c\) is the effective collision frequency, and \(\omega_p\) is the electron plasma frequency,
\begin{equation}
\omega_p
=
\left(
\frac{n_e e^2}{\epsilon_0 m_e}
\right)^{1/2}.
\label{eq:plasma_frequency}
\end{equation}
The real and imaginary parts of the relative permittivity are given by
\begin{equation}
\epsilon_r
=
1-
\frac{\omega_p^2}
{\omega_{\mathrm{0}}^2+\nu_c^2},
\label{eq:epsilon_real}
\end{equation}
and
\begin{equation}
\epsilon_i
=
\frac{\omega_p^2\nu_c}
{\omega_{\mathrm{0}}\left(\omega_{\mathrm{0}}^2+\nu_c^2\right)}.
\label{eq:epsilon_imag}
\end{equation}
The corresponding complex refractive index is
\begin{equation}
\tilde{N}
=
\sqrt{\epsilon_r + i\epsilon_i},
\label{eq:complex_index}
\end{equation} 
For phase measurements, the real part of the refractive index can be expressed as
\begin{equation}
N_r
=
\left[
\frac{
\sqrt{\epsilon_r^2+\epsilon_i^2}
+
\epsilon_r
}{2}
\right]^{1/2}.
\label{eq:real_refractive_index}
\end{equation}
The plasma-induced phase shift is then given by
\begin{equation}
\Delta\phi
=
\frac{2\pi}{\lambda_{\mathrm{0}}}
\int
\left[
N_r-1
\right]dl,
\label{eq:phase_collision}
\end{equation}
where \(\lambda_{\mathrm{0}}=266~\mathrm{nm}\) is the probe wavelength.

To examine the effect of the collision frequency in a dimensionless form, we define
\begin{equation}
X
\equiv
\frac{\nu_c}{\omega_{\mathrm{0}}},
\qquad
\eta
\equiv
\frac{n_e}{n_c},
\label{eq:dimensionless_parameters}
\end{equation}
where \(n_c\) is the critical density of the probe beam,
\begin{equation}
n_c
=
\frac{\epsilon_0 m_e\omega_{\mathrm{0}}^2}{e^2}.
\label{eq:critical_density}
\end{equation}
Using these definitions, equations~\eqref{eq:epsilon_real} and \eqref{eq:epsilon_imag} can be rewritten as
\begin{equation}
\epsilon_r
=
1-
\frac{\eta}{1+X^2},
\label{eq:epsilon_real_dimensionless}
\end{equation}
and
\begin{equation}
\epsilon_i
=
\eta
\frac{X}{1+X^2}.
\label{eq:epsilon_imag_dimensionless}
\end{equation}

For the 10 ns phase-shift map, the maximum measured phase shift was approximately \(-0.53\pi\). 
To estimate the possible density correction caused by collisions, we considered a simplified uniform plasma slab with an effective path length of \(L_p=1~\mathrm{mm}\), corresponding to the approximate plasma diameter observed in the experiment. 
Using the collisionless approximation, the line-of-sight-averaged electron density is
\begin{equation}
\bar{n}_{e,0}
=
-
\frac{\lambda_{\mathrm{0}} n_c}{\pi L_p}
\Delta\phi,
\label{eq:collisionless_average_density}
\end{equation}
where the negative sign reflects the fact that the plasma refractive index is smaller than unity. 
For \(\Delta\phi=-0.53\pi\), this gives
\begin{equation}
\bar{n}_{e,0}
\simeq
2.1\times10^{18}~\mathrm{cm^{-3}},
\end{equation}
as indicated by the red solid line in figure~\ref{fig:fidelity}(c).

The collision-corrected density was calculated by solving equation~\eqref{eq:phase_collision} with the complex dielectric response in equations~\eqref{eq:epsilon_real_dimensionless} and \eqref{eq:epsilon_imag_dimensionless}. 
The ratio \(X=\nu_c/\omega_{\mathrm{0}}\) was varied from \(10^{-3}\) to 1, where \(X=1\) corresponds to the extreme case in which the collision frequency is equal to the probe angular frequency. 
The resulting collision-corrected electron density is shown by the black solid line in figure~\ref{fig:fidelity}(c).

The calculated density remains almost identical to the collisionless estimate for \(X \lesssim 0.1\). 
A noticeable difference appears when the collision frequency becomes larger than approximately \(0.1\omega_{\mathrm{0}}\). 
Even in the extreme case of \(X=1\), however, the estimated line-of-sight-averaged electron density is approximately
\begin{equation}
\bar{n}_{e}(X=1)
\simeq
4.2\times10^{18}~\mathrm{cm^{-3}},
\end{equation}
which is only about twice the collisionless value and does not differ by more than an order of magnitude.
Therefore, although finite collision frequency can lead to an underestimation of the electron density when the collisionless assumption is used, the present analysis shows that this effect does not invalidate the order of magnitude of the electron density inferred for the dense helium plasma studied here.

\section{Summary}

We developed a nanosecond-resolved 266 nm Mach-Zehnder interferometer and applied it to electron-density measurements of dense helium laser-produced plasmas generated in a high-pressure SCF medium. 
The use of a 266 nm UV probe provides a critical density on the order of \(10^{22}~\mathrm{cm^{-3}}\), thereby extending the accessible density range compared with VIS-IR interferometry and reducing the possibility of probe cutoff under the present helium plasma conditions.
To improve the reliability of the phase reconstruction, plasma-arm-only and reference-arm-only images were acquired and used to correct the raw-interferogram fringe patterns. 
This fringe-correction procedure compensated for the spatial intensity imbalance and improved the effective fringe visibility in the plasma region. 
The corrected interferograms were then processed using a 2D FFT to reconstruct the 2D phase-shift maps. 
Using the plasma dispersion relation, these phase-shift maps were converted into line-integrated electron-density distributions.
Assuming cylindrical symmetry of the laser-produced plasma about the \(z\)-axis, Abel inversion was applied to the plasma column showing the largest line-integrated electron density. 
The maximum local electron density was estimated to be \(2.5\times10^{18}~\mathrm{cm^{-3}}\). 
This result demonstrates that the present 266 nm UV Mach-Zehnder interferometer can provide quantitative, nanosecond-resolved electron-density information for dense plasmas generated in a high-pressure supercritical-fluid medium.

The fidelity of the interferometry was evaluated by considering three possible sources of error: probe-beam absorption, probe-beam refraction, and the effect of finite collision frequency in dense plasmas. 
For probe absorption, the optical depth was estimated using the plasma diameter and the free-free absorption coefficient. 
The estimated optical depth satisfied \(\alpha_{ff} L_p < 1\), indicating that probe absorption was not strong enough to invalidate the measurement. 
For probe refraction, the optical geometry of the interferometer was used to estimate the maximum allowed transverse electron-density variation condition for high fringe contrast. 
The calculation showed that the interferometer can retain sufficient fringe contrast up to an electron-density variation of approximately \(1.16\times10^{20}~\mathrm{cm^{-3}}\). 
Finally, we investigated the effect of finite collision frequency on the inferred electron density. 
The analysis showed that neglecting the collision frequency can lead to an underestimation of the electron density for a given phase shift when the ratio between the collision frequency and the probe angular frequency becomes significant.
However, even when the collision frequency is assumed to be the same as the probe angular frequency, the electron density does not differ from the collisionless estimate by more than an order of magnitude. 
These results support the validity of the approximations used to infer the electron density.

Time-resolved interferometry of dense plasmas is essential for understanding non-ideal plasma states, because electron density is a fundamental parameter required to evaluate plasma coupling, degeneracy, transport, and optical response. 
The present work demonstrates that 266 nm UV Mach-Zehnder interferometry can provide a quantitatively reliable diagnostic platform for nanosecond-scale electron-density measurements in dense laser-produced plasmas. 
In future work, this interferometric diagnostic will be combined with plasma imaging and optical emission spectroscopy so that the plasma geometry, temperature, and electron density can be measured. 
Additional schlieren imaging will also be implemented to cross-check the interferometric results and to qualitatively assess the shock wave generated by the laser-produced plasma.
The integrated diagnostic system will provide a useful experimental basis for studying dense plasmas generated by coupling high-power lasers with high-pressure supercritical-fluid media.

\begin{acknowledgments}
This work was supported by the National Research Foundation of Korea (NRF) grant funded by the Ministry of Science and ICT (RS-2024-00349684) and partially supported by Glocal University 30 Project funded by the Ministry of Education.
\end{acknowledgments}

\section*{Author Declarations}

\subsection*{Conflict of Interest}
The authors have no conflicts to disclose.

\subsection*{Author Contributions}
Kyusang Cho: Conceptualization (equal); Data curation (lead); Formal analysis (equal); Investigation (equal); Methodology (equal); Visualization (lead); Writing - original draft (lead).
Juho Lee: Conceptualization (equal); Data curation (supporting); Formal analysis (equal); Investigation (equal); Methodology (equal); Visualization (supporting).
Gunsu Yun: Conceptualization (equal); Funding acquisition (lead); Project administration (lead); Supervision (lead).

\section*{Data Availability}
The data that support the findings of this study are available from the corresponding author upon reasonable request.

\bibliography{references}

\end{document}